\documentclass[12pt]{article}

\usepackage{fancyhdr,a4}
\usepackage{graphics}
\usepackage{epsfig,epsf}
\usepackage[hang,small,bf]{caption}
\usepackage{amssymb}
\usepackage{supertabular}

\begin{document}

\def\dsdt{$\frac{d\sigma}{dt}$}
\def\beqn{\begin{eqnarray}}
\def\eeqn{\end{eqnarray}}
\def\barr{\begin{array}}
\def\earr{\end{array}}
\def\btab{\begin{tabular}}
\def\etab{\end{tabular}}
\def\bite{\begin{itemize}}
\def\eite{\end{itemize}}
\def\bcen{\begin{center}}
\def\ecen{\end{center}}

\def\eq{\begin{equation}}
\def\ee{\end{equation}}
\def\eqa{\begin{eqnarray}}
\def\eea{\end{eqnarray}}
\def\nn{\nonumber}
\def\psmu{P^{\prime \mu}}
\def\psnu{P^{\prime \nu}}
\def\ksmu{K^{\prime \mu}}
\def\pss{P^{\prime \hspace{0.05cm}2}}
\def\psf{P^{\prime \hspace{0.05cm}4}}
\def\kdagger{K\hspace{-0.3cm}/}
\def\ndagger{N\hspace{-0.3cm}/}
\def\psdagger{p'\hspace{-0.28cm}/}
\def\epssdagger{\varepsilon'\hspace{-0.28cm}/}
\def\epsdagger{\varepsilon\hspace{-0.18cm}/}
\def\pdagger{p\hspace{-0.18cm}/}
\def\xidagger{\xi\hspace{-0.18cm}/}
\def\qsdagger{q'\hspace{-0.3cm}/}
\def\qdagger{q\hspace{-0.2cm}/}
\def\keldagger{k\hspace{-0.2cm}/}
\def\ksdagger{k'\hspace{-0.3cm}/}
\def\q2dagger{q_2\hspace{-0.35cm}/\;}
\def\qqs{q\!\cdot\!q'}
\def\lls{l\!\cdot\!l'}
\def\lp{l\!\cdot\!p}
\def\lps{l\!\cdot\!p'}
\def\lsp{l'\!\cdot\!p}
\def\lsps{l'\!\cdot\!p'}
\def\lqs{l\!\cdot\!q'}
\def\pps{p\!\cdot\!p'}
\def\psqs{p'\!\cdot\!q'}
\def\epsp{\varepsilon'\!\cdot\!p}
\def\epsps{\varepsilon'\!\cdot\!p'}
\def\epsl{\varepsilon'\!\cdot\!l}
\def\epsls{\varepsilon'\!\cdot\!l'}

\title{Generalized sum rules of the nucleon in the constituent quark model}
\author{M. Gorchtein$^1$, D. Drechsel$^2$, M.M. Giannini$^1$, E. Santopinto$^1$
and L. Tiator$^2$}
\maketitle
$^1$ Universit\`a di Genova, Sezione INFN di Genova, \\
\indent
via Dodecaneso 33, 16146 Genova (Italy)\\
\indent
$^2$ Institut f\"ur Kernphysik, Universit\"at Mainz, \\
\indent
J.-J.-Becherweg 45, 55099 Mainz (Germany)
\abstract{
We study the generalized sum rules and polarizabilities of the nucleon in the
framework of the hypercentral constituent quark model. We include in the
calculation all the well known $3^*$ and $4^*$ resonances and consider all the
generalized sum rules for which there are data available.
To test the model dependence of the calculation, we compare our results
to the results obtained in the harmonic oscillator CQM.
We furthermore confront our results to the model-independent sum rules values
and to the predictions of the phenomenological
MAID model. The CQM calculations provide a good description of most of the
presented generalized sum rules in the intermediate $Q^2$ region (above
$\sim0.2$ GeV$^2$) while they encounter difficulties in describing
these observables at low $Q^2$, where the effects of the pion cloud, not
included in the present calculation, are expected to be important.
} 

\section{Introduction}
\label{intro}

The sum rules for real and virtual Compton scattering are constructed as
energy-weighted integrals over the various contributions to the inclusive cross
section. We list the generalized sum rules in Section \ref{sec:sumrules} and
refer the reader to Refs. \cite{reviewdr},\cite{drechsel} for details of the
derivation. The sum rules serve as a powerful tool to study the nucleon
structure by providing a bridge between the static properties of the nucleon
(such as charge, mass, and magnetic moment) and the dynamical properties (e.g.,
the transition amplitudes to excited states) in a wide range of energy and
momentum transfer $Q^2$. Since a constituent quark model (CQM) provides
predictions for both the static and the dynamic properties, the sum rules make
it possible to test the consistency of the model. Recently, precise
measurements of the generalized sum rules and related observables have become
available in a series of experiments \cite{MAMI}-\cite{JLABneutron}.
Furthermore, the MAID model \cite{DHKT}, \cite{DKL} yields a detailed analysis
of the (mainly) single pion photo- and electroproduction channels in a wide
energy and $Q^2$ range.

In recent years much attention has been devoted to the description of the
internal nucleon structure in terms of constituent quark degrees of
freedom. Besides the now classical Isgur-Karl model \cite{KI},
the Constituent Quark Model has been proposed in quite different approaches:
the Capstick and Isgur model \cite{capstick},
the chiral model \cite{olof}, \cite{ple}, the algebraic $U(7)$-model
\cite{bijker}, and the hypercentral formulation \cite{hccqm1}.

In this work we study the generalized sum rules for the nucleon, briefly
reviewed in Section \ref{sec:sumrules}, within the
hypercentral constituent quark model. In Section \ref{sec:cqm}, we discuss
the main features of this model and obtain the expressions for the
generalized sum rules within the zero-width approximation.
For this calculation, we consistently use the parameters of the model fixed
to the baryonic spectrum as given by previous publications
\cite{hccqm1}-\cite{hccqm5}. In the past, a series of
calculations were performed in various CQM models \cite{gdhcqm, gdh:thomas},
however all of them restricted themselves to the GDH
sum rule only and had their scope in reproducing the value of this sum rule
at $Q^2=0$. At finite values of the momentum transfer, the experimental data
for the first moment of the
proton DIS structure function $g_1$ were compared to the CQM calculation
in Ref.\cite{gdh:thomas}. Thanks to the availability of many new precise
data for different generalized sum rules for both the proton and the neutron,
we now attempt to provide a description of all the nucleon
sum rules using the same model without any new ingredients as compared to the
baryonic spectrum calculation. Since for the moment it is impossible to
account for the $q\bar{q}$ (pionic) degrees of freedom in a model independent
way, such contributions are not included in this work.
In our calculation, we include the 14 well-known resonances ($3^*$ and $4^*$
in the classification of PDG). To test the dependence on the
particular quark model, we also consider the harmonic oscillator type
of the CQM. In Section \ref{sec:results}, we present our predictions
for the generalized sum rules for the proton and neutron and compare them to
the available data. Finally, in Appendix \ref{sec:appendix}, we provide
details of the derivation of the inclusive cross sections in the helicity
representation.

\section{Generalized sum rules}
\label{sec:sumrules}

The inclusive cross section for electron-proton scattering can be written as
follows:
\beqn
\frac{d\sigma}{d\Omega dE'}&=&\Gamma_V
\left[ \sigma_T\,+\,\epsilon\sigma_L
\,-\,hP_z\sqrt{1-\epsilon^2}\sigma_{TT}
\,-\,hP_x\sqrt{2\epsilon(1-\epsilon)}\sigma_{LT} \right]\;,
\label{eq:crosssec}
\eeqn
with the virtual photon flux factor
 \beqn
 \Gamma_V\;=\;\frac{\alpha_{em}}{2\pi^2}\frac{E'}{E}\frac{K}{Q^2}
 \frac{1}{1-\epsilon}\;,
 \eeqn
and the photon polarization
\beqn
\epsilon\;=\;\frac{1}{1+2(1+\nu^2/Q^2)\tan^2(\Theta/2)}\;,
\eeqn
where $E(E')$ denote the initial (final) electron energy, $\nu=E-E'$ the
energy transfer to the target, $\Theta$ the electron scattering angle, and
$Q^2=4EE'\sin^2(\Theta/2)$ the
four-momentum transfer. The virtual photon spectrum normalization factor is
chosen according to Hand's definition,
$K\,=\,K_H\,\equiv\,\nu-\frac{Q^2}{2M}$, with $M$ the
nucleon mass. Furthermore, the polarization in Eq.(\ref{eq:crosssec}) is
described by the helicity of the electron, $h=\pm 1$, while $P_z$ and $P_x$
are the components of the target polarization along the virtual photon
momentum and perpendicularly to it in the leptonic plane. The partial cross
sections are related to the nucleon structure functions,
\beqn
\sigma_T&=&\frac{4\pi^2\alpha_{em}}{MK}\,F_1\;,\nn\\
\sigma_{TT}&=&\frac{4\pi^2\alpha_{em}}{MK}\,(g_1\,-\,\gamma^2g_2)\;,\nn\\
\sigma_L&=&\frac{4\pi^2\alpha_{em}}{K}\,
\left\{
\frac{1+\gamma^2}{\gamma^2}\frac{F_2}{\nu}\,-\,\frac{F_1}{M}
\right\}\;,\nn\\
\sigma_{LT}&=&\frac{4\pi^2\alpha_{em}}{MK}\,\gamma\,(g_1\,+\,g_2)\;,
\eeqn
with $\gamma=Q/\nu$.

The Baldin sum rule \cite{baldin} relates the sum of the electric
polarizability $\alpha$ and the magnetic susceptibility $\beta$ to the
following energy weighted integral over the total photoabsorption cross
section:
\beqn
\alpha(Q^2)\,+\,\beta(Q^2)&=&\frac{1}{2\pi^2}
\int_{\nu_0}^\infty \frac{K(\nu,Q^2)}{\nu}\,
\frac{\sigma_T(\nu, Q^2)}{\nu^2}d\nu\;,
\label{eq:baldinq2}
\eeqn
where $\nu_0$ is the threshold energy for pion production. Another sum rule
expresses the longitudinal polarizability $\alpha_L$ by an integral over the
longitudinal cross section $\sigma_L$ \cite{reviewdr},
\beqn
\alpha_L(Q^2)&=&\frac{1}{2\pi^2}\int_{\nu_0}^\infty \frac{K(\nu,Q^2)}{\nu}\,
\frac{\sigma_L(\nu, Q^2)}{\nu^2}d\nu\;.
\label{eq:alphal}
\eeqn

The forward spin polarizability $\gamma_0$ can be calculated from the helicity
difference $\sigma_{TT}$,
\beqn
\gamma_{TT}(Q^2)&=&\frac{1}{2\pi^2}\int_{\nu_0}^\infty
\frac{K(\nu,Q^2)}{\nu}\,\frac{\sigma_{TT}(\nu, Q^2)}{\nu^3}d\nu\;.
\label{eq:gamma0q2}
\eeqn

Similarly, we obtain the longitudinal-transverse polarizability $\delta_{LT}$
from the integral
\beqn
\delta_{LT}(Q^2)&=&\frac{1}{2\pi^2}\int_{\nu_0}^\infty \frac{K(\nu,Q^2)}{\nu}\,
\frac{\sigma_{LT}(\nu, Q^2)}{Q\nu^2}d\nu\;.
\label{eq:deltaltq2}
\eeqn

Further sum rules are given by generalizations of the GDH sum rule~\cite{gdh},
the purely transverse expression
\beqn
I_{TT}(Q^2)&=&\frac{M^2}{\pi e^2}\int_{\nu_0}^\infty
\frac{K(\nu,Q^2)}{\nu}\,\frac{\sigma_{TT}(\nu, Q^2)}{\nu}d\nu\;,
\label{eq:gdhq2}
\eeqn
and the first moment of the structure function $g_1$ known from deep inelastic
scattering (DIS),
\beqn
I_1^N(Q^2)&=&\frac{2M^2}{Q^2}\int_0^{x_0}g_1^Ndx
\;=\;\frac{M^2}{\pi e^2}\,\int_{\nu_0}^\infty\frac{K}{\nu^2+Q^2}
\left[ \sigma_{TT}\,+\,\frac{Q}{\nu}\sigma_{LT} \right]d\nu \nn\\
&\rightarrow&
\left\{ \begin{array}{c} -\frac{\kappa_N^2}{4},\;Q^2\rightarrow 0\nn\\
\frac{2M^2}{Q^2}\Gamma_1^N,\;Q^2\rightarrow \infty\nn
\end{array} \right.
\eeqn
where $\Gamma_1^N\,=\,\int_0^1dxg_1^N(x,Q^2)$ and $x_0=Q^2/(2M\nu_0)$.

The Bjorken sum rule \cite{bjorken} deals with the isovector combination of
$I_1$,
\beqn
\Gamma_1^p\,-\Gamma_1^n\;=\;\int_0^1 dx\,[g_1^p(x,Q^2)\,-\,g_1^n(x,Q^2)]
\;\rightarrow\;
\frac{1}{6}g_A\;\;{\rm at}\;Q^2\,\rightarrow\,\infty\;,
\label{bjorken}
\eeqn
\noindent
with $g_A=1.26$ the axial coupling constant of the nucleon. This sum rule is
well established both theoretically and experimentally, i.e., at $Q^2=5$ GeV,
\beqn
\left(\Gamma_1^p\,-\,\Gamma_1^n\right)^{exp}
&=&0.176\pm 0.003\pm 0.007 \;\;\;\cite{bjorkenexp}\nn\\
\left(\Gamma_1^p\,-\,\Gamma_1^n\right)^{th}
&=&0.182\pm 0.005 \;\;\;\cite{bjorkentheo}\;,
\eeqn
\noindent
where the theoretical value contains the radiative corrections to
Eq.(\ref{bjorken}).

In a similar way the Burkhardt-Cottingham (BC) sum rule \cite{bc} involves
the first moment of the (DIS) structure function $g_2$. The BC sum rule
states that this moment vanishes if integrated over elastic and inelastic
contributions,
\beqn
\int_0^1g_2dx\;=\;0\;\;{\rm for\;any}\;Q^2\;.
\eeqn

If this relation holds, the inelastic contribution to the integral can be
expressed by the nucleon form factors for all values of $Q^2$,
\beqn
I_2(Q^2)\;=\;\frac{2M^2}{Q^2}\int_0^{x_0}g_2dx
&=&\frac{M^2}{\pi e^2}\,\int_{\nu_0}^\infty
\frac{K}{\nu^2+Q^2} \left[ -\sigma_{TT}\,+\,\frac{\nu}{Q}\sigma_{LT} \right]
d\nu\nn\\
&=&{1\over 4}\frac{G_M(Q^2)(G_M(Q^2)-G_E(Q^2))}{1+Q^2/4M^2}\;.
\eeqn

Furthermore, if we add the integrals $I_1$ and $I_2$, we obtain the purely
longitudinal-transverse expression
\begin{equation}
I_{LT}(Q^2)=\frac{M^2}{\pi e^2}\int_{\nu_0}^\infty
\frac{K(\nu,Q^2)}{\nu}\,\frac{1}{Q}\,\sigma_{LT}(\nu, Q^2)d\nu\;.
\label{eq:i3q2}
\end{equation}

At the real photon point, $Q^2=0$, the values of the sum rules take the
following values \cite{reviewdr}:
\beqn
\alpha^p(0)+\beta^p(0)&=&13.69\cdot 10^{-4}fm^3\;\;\;\cite{bab}\;,\nn\\
\alpha_L(0)&=&0\nn\\
\gamma_{TT}^p(0)&=&-1.01\cdot 10^{-4}fm^4\;\;\;\cite{MAMI}\nn\\
I_{TT}(0)&=&-\frac{\kappa_N^2}{4}\nn\\
I_1(0)&=&-\frac{\kappa_N^2}{4}\nn\\
I_2(0)&=&\frac{\kappa_N(e_N+\kappa_N)}{4}\nn\\
I_{LT}(0)&=&\frac{e_N\kappa_N}{4}\;,
\label{eq:sumrulevalues}
\eeqn
with $e_p=1,\;e_n=0,\;\kappa_p=1.79,\;\kappa_n=-1.91$.

\section{Constituent quark model}
\label{sec:cqm}

In the following we shall review the hypercentral Constituent Quark Model
(HCQM), which has been used for a systematic calculation of various baryon
properties.

The experimental $3^*$ and $4^*$ non strange resonances can be arranged
in  $SU(6)-$multiplets. This means that the quark dynamics
has a dominant $SU(6)-$ invariant part, which accounts for the average
multiplet energies. In the HCQM the potential is assumed to take the form
\cite{hccqm1}
\begin{equation}
\label{eq:pot}
V(x)= -\frac{\tau}{x}~+~\alpha x\,,
\end{equation}
with the hyperradius $x=\sqrt{{\vec{\rho}}\,^2+{\vec{\lambda}}^2}$,
where $\vec{\rho}$ and $\vec{\lambda}$ are the Jacobi coordinates
describing the
internal quark motion. We note that the ``hypercentral'' potential
does not depend on the hyperangle $\xi=arctg(\rho/\lambda)$ .\\
Interactions of the type linear plus Coulomb-like have been used for a
long time in the meson sector, e.g. the Cornell potential. This form has
been supported by recent Lattice QCD calculations \cite{bali}.\\
In the case of baryons a so called hypercentral approximation was
introduced \cite{has}, which amounts to average any two-body
potential for the three quark system over the hyperangle $\xi$.
This approximation works
quite well, especially for the lower part of the spectrum \cite{hca}. In
this respect, the hypercentral potential Eq.(\ref{eq:pot}) can be considered
as the hypercentral approximation of a two-body Coulomb-like plus
linear confining potential.\\

The hypercoulomb term $1/x$ has important
features \cite{hccqm1,sig}: it can be solved analytically and the resulting
form factors have a power-law behaviour, at variance with the widely used
harmonic oscillator. Moreover, the negative parity states are exactly
degenerate with the first positive parity excitation, which provides a good
starting point for the description of the spectrum.\\

The splittings within the multiplets are produced by a perturbative term
breaking the $SU(6)$ symmetry,
which, as a first approximation, can be assumed to be the
standard hyperfine interaction $H_{hyp}$ \cite{KI}.
The three quark hamiltonian for the HCQM is then:
\begin{equation}\label{eq:ham}
H = \frac{p_{\lambda}^2}{2m}+\frac{p_{\rho}^2}{2m}-\frac{\tau}{x}~
+~\alpha x+H_{hyp},
\end{equation}
where $m$ is the quark mass (taken equal to $1/3$ of the nucleon mass).
The strength of the hyperfine interaction is determined such as to
reproduce the $\Delta-N$ mass difference, the two remaining free
parameters are fitted to the spectrum,
which yields the following values:
\begin{equation}\label{eq:par}
\alpha= 1.61~fm^{-2},~~~~\tau=4.59~.
\end{equation}

Keeping these parameters fixed, the model has been applied to calculate
various physical quantities of interest: the photocouplings \cite{hccqm2},
electromagnetic transition amplitudes \cite{hccqm3}, the elastic nucleon
form factors \cite{hccqm4}, in particular the ratio between the electric
and magnetic form factors,
and  recently in a systematic calculation of the longitudinal
electromagnetic form factors for all the $3^*$ and $4^*$
resonances \cite{hcqmlong}.

We also observe that the harmonic oscillator potential, which
is widely used in quark models because of its analytical solution, is in fact
exactly hypercentral,
\beqn
\sum_{i<j}{1\over 2}k(\vec{r}_i\,-\,\vec{r}_j)^2\;=\;{3\over 2}kx^2
\;=\;V_{HO}(x)\;,
\eeqn
and thus can be treated in the hypercentral approach.

For comparison, we shall also show the analytical results within the
constituent quark model with the harmonic oscillator potential (HO) of
Ref.~\cite{KI}, which gives analytical results for the transverse helicity
amplitudes $A_{1/2}$ and $A_{3/2}$. The results for the
longitudinal helicity amplitudes $S_{1/2}$ are constructed in a similar way.
We now relate the partial cross sections to the helicity amplitudes for
resonance excitation. In the following we use the $lab$ frame with the proton
four-momentum $p^\mu\,=\,(M,\vec{0})$, and $q^\mu\,=\,(\nu,0,0,q)$ the
four-momentum of the virtual photon. The photon polarization vectors take the
form:
\beqn
\varepsilon^{(\pm)}&=&\mp\frac{1}{\sqrt{2}}(0,1,\pm i,0) \nn\\
\varepsilon^{(0)}&=&\frac{1}{Q}(q,0,0,\nu)\;.
\eeqn

The spherical components of the (three-vector) current are defined by
$J_{\pm}=\mp (J_x\pm i\, J_y)/\sqrt{2}$ and $J_0 = J_z$. The latter is related to
the charge by  gauge invariance, $q \cdot J= \nu\rho\,-\,qJ_z\,=0$. From these equations
we find
\beqn
\varepsilon^{(\pm)} \cdot J & = & - J_{\pm}\ , \nonumber \\
\varepsilon^{(0)} \cdot J & = & \frac{Q}{\nu}\,J_0 = \frac{Q}{q}\rho\;,
\label{eps}
\eeqn
which leads to a Lorentz invariant description of the transition operators.
The electromagnetic transition helicity amplitudes are then defined by

\beqn
A_{1/2}&=&-\frac{e}{\sqrt{2K}} <R,{1\over 2}|J_+|N,-{1\over 2}>\xi\;, \nn\\
A_{3/2}&=&-\frac{e}{\sqrt{2K}} <R,{3\over 2}|J_+|N,{1\over 2}>\xi\;, \nn\\
S_{1/2}&=&-\frac{e}{\sqrt{2K}} <R,{1\over 2}|\rho|N,{1\over 2}>\xi\;,
\label{eq:helampldefinition}
\eeqn where the 3rd component of the nucleon and resonance spins are explicitly
indicated. The phases $\xi$ depend on the strong decay matrix elements, which
have to be individually calculated in the respective model. In comparing with
Eq.~(\ref{eps}), we note that the $S_{1/2}$ amplitude is finite at the real
photon point ($Q^2=0$). However, only the product $\frac{Q}{q}\,S_{1/2}$
transforms as a Lorentz scalar. A single resonance contribution to the
transverse helicity cross sections is given by~\cite{giannini}

\beqn \sigma_{\Lambda}\;=\;\frac{4M}{M_R\Gamma_R}|A_{\Lambda}|^2\;,\;\;\;\;
\Lambda\,=\,{1\over 2}\;{\rm or}\; {3\over 2}\;,
\label{sigma_Lambda}
\eeqn
with $M_R$ and $\Gamma_R$ the mass and the full width of the resonance.
The total photoabsorption cross section $\sigma_{T}$ and the helicity
difference $\sigma_{TT}$ are then obtained by
\beqn
\sigma_{T}&=&\frac{\sigma_{1/2}+\sigma_{3/2}}{2} \;=\;\frac{2M}{M_R\Gamma_R}
\left\{|A_{1/2}|^2\,+\,|A_{3/2}|^2\right\}\ , \nn \\
\sigma_{TT}&=&\frac{\sigma_{1/2}-\sigma_{3/2}}{2} \;=\;\frac{2M}{M_R\Gamma_R}
\left\{|A_{1/2}|^2\,-\,|A_{3/2}|^2\right\}\ . \eeqn

The helicity amplitudes for the transitions $\gamma^*N\rightarrow R$ can be
directly calculated within the CQM at the resonance position in the limit of
a sharp resonance. In this limit, the energy dependence of the cross section
is described by a $\delta$-function. In a more realistic scenario we describe
a resonance with a finite width $\Gamma_R$ and a resonance energy $M_R$, such
that the real part of the amplitude vanishes exactly at resonance,
\beqn
A_{\Lambda}(W)\;=\;\frac{\Gamma_R/2}{W-M_R-i\Gamma_R/2}\,A_{\Lambda}^0\;,
\label{A_Lambda}
\eeqn where $W = \sqrt{2M\nu + M^2 - Q^2}$ is the total $c.m.$ energy and
$A_{\Lambda}^0$ the helicity amplitude as calculated in the CQM. We note that
these CQM helicity amplitudes can be defined to be real by choice of the
model-dependent phases $\xi$ introduced in Eq.(\ref{eq:helampldefinition}), and
therefore we shall treat these amplitudes as real numbers in the following
equations. Using equations (\ref{sigma_Lambda}) and (\ref{A_Lambda}), we obtain
the cross section in the zero-width approximation, in the limit
$\Gamma_R\rightarrow 0$,
 \beqn \sigma_{\Lambda}^0(W)\;=\;2\pi\delta(W-M_R)
\frac{M}{M_R} (A_{\Lambda}^0)^2 \ . \eeqn

Expressed as function of the photon $lab$ energy $\nu$, the zero-width cross
section takes the form \beqn \sigma_{\Lambda}^0(\nu) &=&
2\pi\delta(\nu-\nu_R)(A_{\Lambda}^0)^2\ , \qquad \Lambda = \frac{1}{2}\
{\rm{and}}\ \frac{3}{2}
\nn\\
\sigma_L^0(\nu) &=& 2\pi\delta(\nu-\nu_R) \left(\frac{Q}{q_R}\right)^2
(S_{1/2}^0)^2\;,
\nn \\
\sigma_{LT}^0 &=& -\sqrt{2}\pi\delta(\nu-\nu_R) \frac{Q}{q_R} S_{1/2}^0
A_{1/2}^0\;.
\label{eq:zerowidthcs}
\eeqn

Since the sign of the longitudinal-transverse interference term is not uniquely
defined in the literature, we address the reader to Appendix~A for a derivation
of the inclusive cross section in the helicity formalism. We are now in a
position to express the sum rules of the previous section in terms of the
helicity amplitudes in the zero-width approximation:

\beqn
&&I_{TT}(Q^2) = \frac{M^2}{e^2}\sum_R \frac{K}{\nu_R^2}\,
\left[(A_{R{1\over2}}^0)^2-(A_{R{3\over2}}^0)^2\right], \\
&&\nn\\
&&I_1(Q^2) = \frac{M^2}{e^2}\,\sum_R\frac{K}{q_R^2} \left\{ (A_{R{1\over
2}}^0)^2\,-\,(A_{R{3\over 2}}^0)^2 \,-\,\frac{\sqrt{2}Q^2}
{\nu_Rq_R}S_{R{1\over 2}}^0 A_{R{1\over 2}}^0 \right\},\qquad \\
&&\nn\\
&&I_2(Q^2) = \frac{M^2}{e^2}\,\sum_R\frac{K}{q_R^2} \left\{(A_{R{3\over
2}}^0)^2\,-\,(A_{R{1\over 2}}^0)^2 \,-\,\sqrt{2} \frac{\nu_R}{q_R}S_{R{1\over
2}}^0
A_{R{1\over 2}}^0 \right\}, \\
&&\nn\\
 &&I_{LT}(Q^2) = -\frac{\sqrt{2}M^2}{e^2}\sum_R \frac{K}{\nu_Rq_R}\,
S_{R{1\over 2}}^0 A_{R{1\over 2}}^0, \\
&&\nn\\
&&\alpha(Q^2)\,+\,\beta(Q^2) = \frac{1}{2\pi}\, \sum_R\frac{K}{\nu_R^3}
\left\{(A_{R{1\over 2}}^0)^2+(A_{R{3\over
2}}^0)^2\right\},\\
&&\nn\\
&&\alpha_L(Q^2) = \frac{1}{\pi}\, \sum_R\frac{K}{\nu_R^3} \frac{Q^2}{q_R^2}
(S_{R{1\over2}}^0)^2 ,\\
&&\nn\\
&&\gamma_{TT}(Q^2) = \frac{1}{2\pi}\, \sum_R\frac{K} {\nu_R^4}
\left\{(A_{R{1\over 2}}^0 )^2-(A_{R{3\over
2}}^0)^2\right\} ,\\
&&\nn\\
&&\delta_{LT}(Q^2) = -\frac{1}{\sqrt{2}\pi}\sum_R \frac{K}
{\nu_R^3q_R}\,S_{R{1\over 2}}^0  A_{R{1\over 2}}^0\,,
\eeqn
where $K=K(\nu_R,Q^2)\,$.

\section{Results}
\label{sec:results}
In this section we present our results for the generalized nucleon sum
rules obtained with the following 14 resonances (3$^*$ and 4$^*$ in the PDG
classification): $P_{33}(1232)$, $P_{11}(1440)$, $S_{11}(1535)$,
$D_{13}(1520)$, $S_{31}(1620)$, $S_{11}(1650)$, $D_{15}(1675)$, $F_{15}(1680)$,
$P_{11}(1710)$, $D_{33}(1700)$, $P_{13}(1720)$, $D_{13}(1700)$,
$F_{35}(1905)$, $F_{37}(1950)$.

\subsection{Generalized GDH integrals}

\begin{figure}[h]
\vspace{-0.1cm}
\begin{center}
\epsfxsize=27pc
\centerline{\epsffile{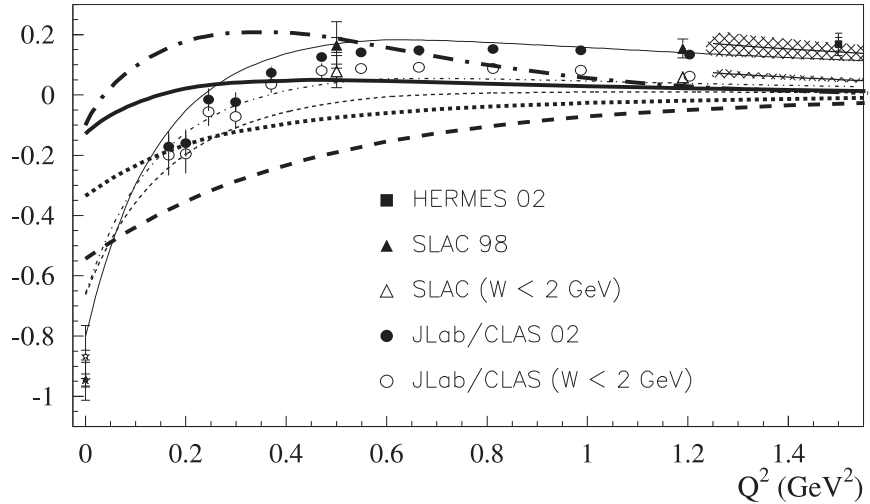}}
\end{center}
\vspace{-1.2cm} \caption{Predictions of the HCQM and HO models for the integral
$I_1^p$ as compared to the MAID results. The HCQM results are shown for the
$\Delta(1232)$ only (thick dashed line), $\Delta(1232)$ plus negative parity
resonances (thick dotted line), and including all 3$^*$ and 4$^*$ resonances
(thick solid line). For the HO model, we only show the full result (thick
dash-dotted line). The thin dashed line corresponds to MAID (one-pion
contribution only). The thin dash-dotted line corresponds to MAID with one and
two-pion contributions. The thin solid line shows the phenomenological
parametrization of Ref. \cite{fit:i1p}. The solid circles (Ref. \cite{JLAB}),
solid triangles (Ref. \cite{SLAC}), and solid squares (Ref. \cite{HERMES})
correspond to the evaluation of the integral over the full energy range, while
the open circles (Ref. \cite{JLAB}) and open triangles (Ref. \cite{SLAC})
include the resonance region only ($W<2$~GeV). For the JLab data \cite{JLAB},
only the statistical errors are shown. The solid lines with shaded bands
starting at 1.25 GeV$^2$ represent the evaluation with the data for the
structure function $g_1$, integrated over the full energy range (upper band)
and over the resonance region only ($W<2$~GeV, lower band), including the
corresponding error estimates \cite{blum}.}
\label{fig:i1p}
\end{figure}

We start with the results for the integrals $I_1^{p}$ and $I_{TT}^{n}$, for
which experimental data are available. In Fig.~\ref{fig:i1p}, we show the
predictions of the two CQMs for $I_1^{p}(Q^2)$ together with the MAID results,
and confront them to the experimental data of Refs. \cite{SLAC} - \cite{JLAB}.

\clearpage

\begin{figure}[h]
\begin{center}
\epsfxsize=25pc
\centerline{\epsffile{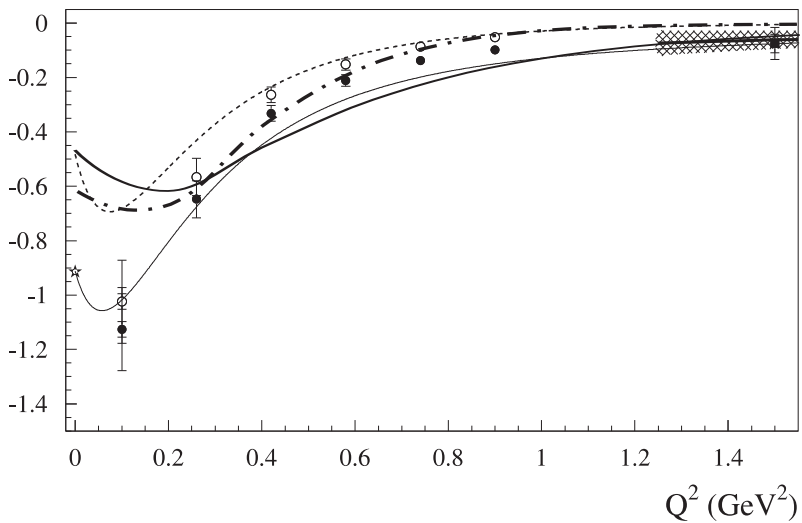}}
\end{center}
\vspace{-0.5cm}
\caption{The CQM predictions for the GDH integral on the neutron, $I_{TT}^n$.
The data points are from Refs. \cite{HERMES} (solid squares) and
\cite{IAnJLAB}
with the nuclear corrections included (solid circles)~\cite{iancorr} and
without these corrections (open circles). The curves represent the results of
HCQM (thick solid),  HO (dashed-dotted), MAID (dashed), and the
parametrization of \cite{fit:i1p} (thin solid), fitted to
the sum rule value at $Q^2=0$ (solid star).}
\label{fig:ian}
\end{figure}

In the range of momentum transfer $0.3<Q^2<0.6$ GeV$^2$,
the HCQM results are compatible with the
experimental data evaluated over the ``resonance'' region ($W\leq 2$ GeV) but
show not enough structure in $Q^2$ to reproduce the data over the full range
of the momentum transfer shown.
However, we find a notable agreement between the HCQM and MAID with one and
two-pion contributions
starting from $Q^2=0.5$ GeV$^2$, where the effects of the pion cloud, included
in MAID but absent in the quark model, are
less important than at low momentum transfers.
At $Q^2=0$, however, both CQ models fail to reproduce the sum rule value.
We shall come back to this point later. \\
Next we present the results for the purely
transverse neutron integral $I_{TT}^{n}$ in Fig.~\ref{fig:ian}.

Again, the two quark models underestimate the strength in
the low $Q^2$ region. Since the effect of the
negative parity resonances is nearly zero in the neutron case, the only
significant deviation from the pure $\Delta(1232)$ contribution comes from
the Roper. Due to characteristic Gaussian form factors, the HO
model is able to reproduce the data only up to $Q^2\approx 1$~GeV$^2$, but
falls short of the data beyond this region. On the contrary, the HCQM
prediction decreases significantly slower with increasing $Q^2$,
as compared to the HO model. However, for $Q^2>0.4$ GeV$^2$ it follows the
phenomenological fit of Ref. \cite{fit:i1p} and
is in very good agreement with the evaluation of the $I_{TT}^n$ integral
with the DIS data and the data point from Ref. \cite{HERMES}.\\

We conclude that the presented CQM does not provide a satisfactory description
of the GDH sum rule at $Q^2=0$. The reason for this is the fact that the
strength at low $Q^2$ is underestimated for the $\Delta(1232)$ case, which
gives the dominant contribution, and similarly for most of the higher
resonances. This indicates the importance of other contributions, such as the
pion cloud effects, not only for the background but for the $N+\gamma\to N^*$
transitions as well. In this respect, it is interesting to note that for
$I_1^p$, the HCQM and MAID results coincide over a large range of intermediate
$Q^2$ values, while they differ at low momentum transfer values, where the pion
loop effects are important.

Although the integral $I_1$ contains a contribution of the
longitudinal-transverse
term, this latter vanishes at the real photon point, and therefore the
disagreement at
$Q^2=0$ is due to the purely transverse resonance contributions.
We can learn more about this fact
by studying another integral containing the helicity cross section
difference, i.e. the forward spin polarizability $\gamma_{TT}$.

\subsection{Forward spin polarizability}

\begin{figure}[h]
\begin{center}
\epsfxsize=22pc
\centerline{\epsffile{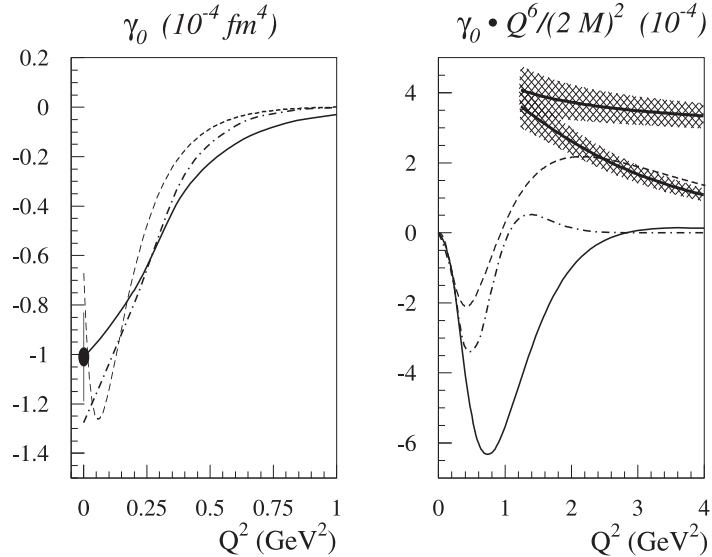}}
\end{center}
\vspace{-1cm}
\caption{Predictions for the generalized proton forward spin polarizability
$\gamma_{TT}$ in units of $10^{-4}fm^4$ (left panel) and
$\gamma_{TT}\cdot(Q^6/(2M)^2)$ in units
of $10^{-4}$ (right panel) within the HCQM (solid lines) and the HO
(dashed-dotted lines) predictions
in comparison with the MAID results for the one-pion contribution only
\cite{DKL}
(dashed lines). The data point is from Ref.~\cite{MAMI}. The shaded bands in
the right
panel show the evaluation of Eq.(\ref{eq:gamma0dis}) with the DIS results for
the
structure functions $g_1$ and $g_2$, integrated over the full energy range
(upper band)
and over the resonance region only ($W<2$~GeV, lower band).  }
\label{fig:gamma0}
\end{figure}

Figure~\ref{fig:gamma0} shows the CQM results and the MAID predictions for the
generalized forward spin polarizability $\gamma_{TT}$ of the proton. The left
panel
shows this observable directly, while the right panel includes an
appropriate factor to make contact with the moments of $g_1$ and $g_2$,
\beqn
\frac{Q^6}{4M^2}\gamma_{TT}\;=\;
\frac{e^2}{\pi}\int_0^{x_0}x^2(g_1\,-\,\frac{4M^2}{Q^2}x^2g_2)dx\;.
\label{eq:gamma0dis}
\eeqn

In this sum rule, there are two more powers of the energy in the denominator
than in the GDH integral, such that the low lying resonances are more
emphasized as compared to the higher resonances than in the case of the GDH. In
fact, it is an easy excercise to prove that if the GDH sum rule is described by
the $\Delta(1232)$ alone, this would result in a value of
$\gamma_{TT}(GDH\equiv\Delta)\approx -1.73\cdot10^{-4}$ fm$^4$, which is much
below the experimental data, see Eq.(\ref{eq:sumrulevalues}). Within a
phenomenological analysis like MAID, one finds that the rather small sum rule
value for $\gamma_{TT}$ comes about due to a destructive interference of large
contributions of the $\Delta(1232)$ resonance and s-wave pion production near
threshold. Of course, the pion cloud also contributes to the GDH integral, but
with a much smaller impact due to the different energy weighting. As can be
seen from Fig.~\ref{fig:gamma0}, both quark models provide fairly good
descriptions for $\gamma_{TT}$ at $Q^2=0$. The absence of the pion threshold
production in the CQM is compensated by the destructive interference effects of
higher resonances and a lack of strength in the resonance region, which can
also be ascribed to a lack of the pion cloud in these transitions.

\subsection{Baldin sum rule}
We proceed with the Baldin sum rule which puts a constraint on the sum of the contributions from
the $J={1\over 2}$ and $J={3\over 2}$ channels instead of their difference, as in the case of the sum rules discussed before.

\begin{figure}[h]
\begin{center}
\epsfxsize=22pc
\centerline{\epsffile{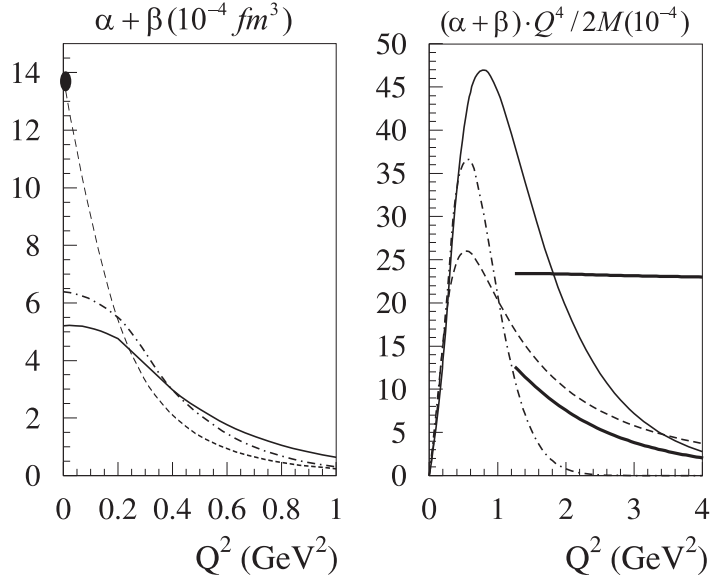}}
\end{center}
\vspace{-1cm}
\caption{Predictions for the generalized proton polarizability
$(\alpha+\beta)$ in units
of $10^{-4}fm^3$ (left panel) and $(\alpha+\beta)\cdot(Q^4/2M)$
in units of $10^{-4}$
(right panel) within HCQM (solid lines) and HO (dashed-dotted lines)
in comparison with the MAID results including the two-pion contribution
\cite{DKL}
(dashed lines). The solid circle corresponds to the evaluation of the Baldin
sum rule
by Ref.~\cite{bab}. The thick solid lines in the right panel are obtained by
evaluating
Eq.(\ref{eq:baldindis}) with the DIS structure function $F_1^p$, integrated
over the
full spectrum (upper curve) and over the resonance region only (lower curve,
$W\leq 2$~GeV).}
\label{fig:baldin}
\end{figure}

In Fig.~\ref{fig:baldin}, we compare
the CQM results with the MAID prediction for the Baldin sum rule of the
proton. While
the left panel displays $\alpha + \beta$ directly, the right panel shows
this observable
with an appropriate factor to obtain the lowest moment of the DIS structure
function $F_1$,
\beqn
\frac{Q^4}{2M}(\alpha\,+\,\beta)
\;=\;\frac{e^2}{\pi}\int_0^{x_0}xF_1(x,Q^2)dx\;.
\label{eq:baldindis}
\eeqn

At $Q^2=0$, both CQM models lack considerable strength, which again can be
explained by the
absence of the pion cloud contribution in these models. Due to the energy
weighting of
the integrals with $1/\nu^2$, the long-range pion contribution is dominant
at low $Q^2$
but decreases with $Q^2$ substantially faster than the resonance contribution.
At larger values of $Q^2$, the HO results are practically zero above $Q^2=2$
(GeV/c)$^2$, due to the strong Gaussian form factor of the HO potential, while
the HCQM curves agree reasonably well with the ``resonant DIS'' and MAID
results. However, the gap between those predictions and the full integral
evaluated with the DIS structure function $F_1$ indicates that all the models
lack important contributions in this $Q^2$ range due to the neglect of the
contributions above $W=2$~GeV. \\

Combining the information from the discussed integrals, we find that the
description of the GDH sum rule with the $\Delta(1232)$ resonance alone, which
has been widely used in the literature \cite{gdhcqm}, turns out to be
inconsistent for the other sum rules, since it leads to an overestimate of
$\gamma_{TT}$ by about a factor of 2 and an even more dramatic underestimation
of the Baldin sum rule. In order to correctly reproduce all the sum rules for
transverse photons at $Q^2=0$ , one needs a contribution below the
$\Delta(1232)$ resonance, which affects the various sum rules in a different
way, because of the respective weighting factors ($1/\nu$, $/\nu^2$ and
$1/\nu^3$) involved. Unfortunately, at present no formalism exists which would
allow for an inclusion of such $q\bar{q}$ (``pionic'') effects in a model
independent way. However, at higher values of the momentum transfer, such
mechanisms become less important and as a result, the HCQM is able to reproduce
the experimental data in the range $0.2$ GeV$^2$ $<\,Q^2\,<$ $1.5$ GeV$^2$. The
applicability of the CQM with the HO nucleon wave functions is however
restricted to momentum transfers below $1$ GeV$^2$, because of its Gaussian
form factors.

\subsection{Bjorken sum rule}
\begin{figure}[h]
\begin{center}
\epsfxsize=25pc
\centerline{\epsffile{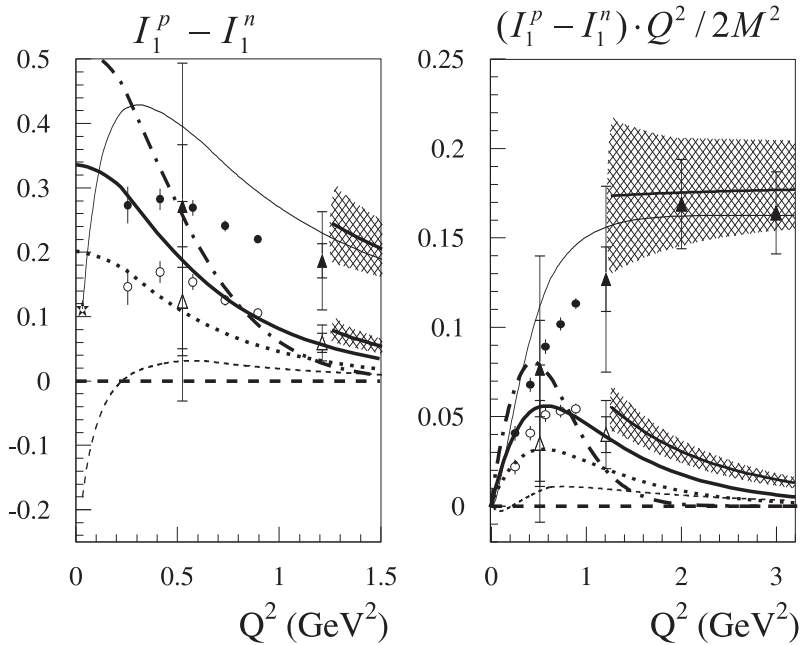}}
\end{center}
\vspace{-0.5cm} \caption{The isovector integral $I_1^p\,-\,I_1^n$ as calculated
in the HCQM and HO models in comparison with the MAID results and the Bjorken
sum rule. The solid and open triangles are data from Ref. \cite{SLAC}. The
solid and open circles represent the combined proton data from \cite{JLAB} and
neutron data from \cite{JLABneutron}, with only statistical errors shown. The
sum rule value is given by the star at $Q^2=0$, and the thin solid line shows
the parametrization of Ref. \cite{fit:i1p}. Further notation as in Fig.
\ref{fig:i1p}.}
\label{fig:i1pmn}
\end{figure}

In Fig. \ref{fig:i1pmn}, we study the isovector combination of $I_1$, which
is fixed by
the Bjorken sum rule for $Q^2\to\infty$. Unlike the MAID model with the
one-pion contribution only, which in this case gives the wrong sign of
the sum rule, the two CQM models predict the right sign of this sum rule but
overestimate its value, which is of course due to the failure to
reproduce the GDH sum rule. Clearly, the isovector
integral is only sensitive to the $N^*$ resonances and not to the $\Delta$
resonances, which contribute equally for proton and neutron.
At $0.5$ GeV$^2<Q^2<1.2$ GeV$^2$, the HCQM model reproduces the
``resonant'' SLAC and JLab data nicely, while the HO model drops too fast due
to its Gaussian form factors.

\subsection{Burkhardt-Cottingham sum rule}

\begin{figure}[h]
\begin{center}
\vspace{-1.cm}
\epsfxsize=30pc
\centerline{\epsffile{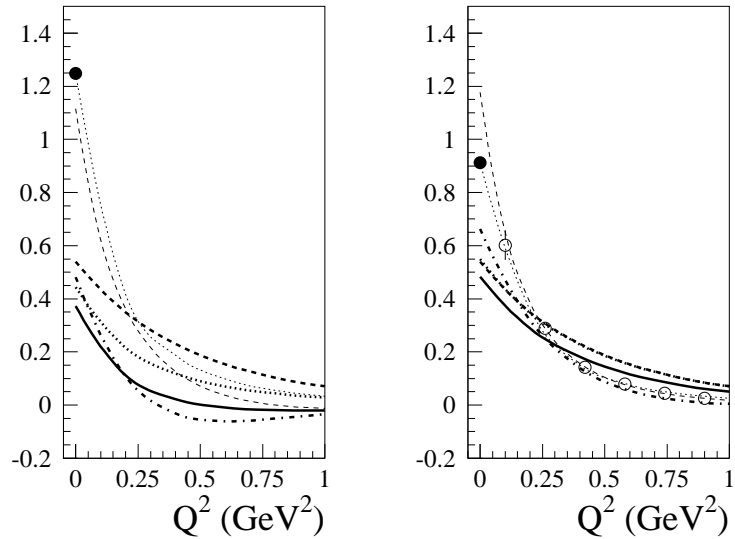}}
\end{center}
\vspace{-1.5cm} \caption{Results for the integral $I_2$ for proton (left panel)
and neutron (right panel). The Burkhardt-Cottingham sum rule corresponds to the
solid circles at $Q^2=0$ and the thin dotted lines. The data are from
\cite{JLABneutron} (open circles) with only the statistical errors shown. 
Further notation as in Fig. \ref{fig:i1p}.}
\label{fig:i2}
\end{figure}

The Burkhardt-Cottingham sum rule connects the $I_2$ integral over the
excitation spectrum to the nucleon form factors for any value of $Q^2$. In Fig.
\ref{fig:i2}, we compare the predictions of the HCQM and HO models for this
integral with the results of MAID, and confront them with the sum rule. For the
proton, MAID agrees within 10$\%$ accuracy with the sum rule already with
one-pion contribution only, however from $Q^2\sim0.2$ GeV$^2$ it starts to
deviate stronger. The two quark models agree over the shown range of $Q^2$,
which implies that the first moment of $g_2^p$ depends on the particular quark
model only weakly. However, both quark models underestimate the sum rule
significantly. In the case of the neutron, new experimental data on the $^3$He
target are available \cite{JLABneutron}. From the comparison to these data, we
find a very good agreement of the HO results with these data and with MAID
starting from 0.25 GeV$^2$. HCQM does agree qualitatively with the data over
the same range as well but lacks structure in $Q^2$ for this observable.

\subsection{Generalized $I_{LT}$ integral}
This sum rule deals with the sum of the first moments of the DIS structure
functions $g_1$ and $g_2$,
\beqn
I_{LT}\;\equiv\;I_1\,+\,I_2\;=\;\int_0^{x_0}(g_1(x,Q^2)+g_2(x,Q^2))dx\;,
\eeqn
and depends on the longitudinal-transverse interference term only.
If the BC and the GDH sum rules hold, this integral is fixed at the real
photon point,
\beqn
I_{LT}(0)\;=\;\frac{e_N\kappa_N}{4}\;.
\label{I3}
\eeqn

\begin{figure}[h]
\vspace{-1cm}
\begin{center}
\epsfxsize=30pc
\centerline{\epsffile{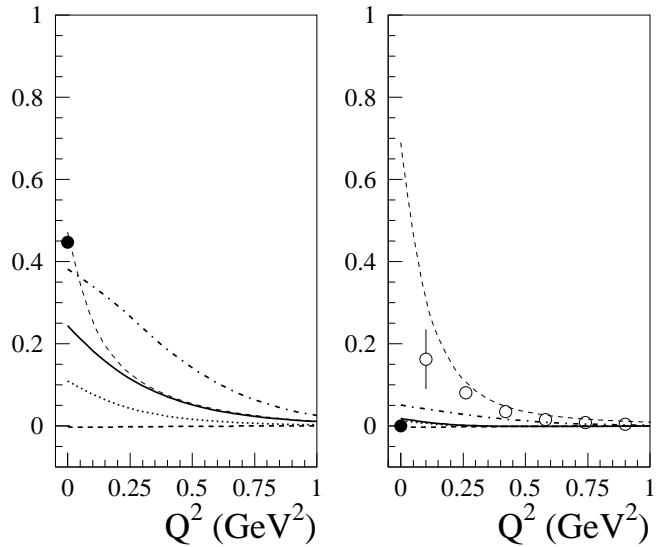}}
\end{center}
\vspace{-1.5 cm}
\caption{The integrals $I_{LT}(Q^2)$ for proton (left panel) and neutron 
(right panel)
as calculated with the HCQM (solid lines) and the HO (dashed-dotted lines) 
models in
comparison with the MAID results for the one-pion contribution (dashed lines). 
The solid circles represent the sum rule values. The data are from
\cite{JLABneutron} (open circles) with only the statistical errors shown. 
Further notation as in Fig. \ref{fig:i1p}.}
\label{fig:i3}
\end{figure}

Our results for $I_{LT}$ are shown in Fig. \ref{fig:i3} for both proton and
neutron. 

In the case of the proton (left panel), the HO and MAID models agree
reasonably well with Eq.(\ref{I3}), while the HCQM model
lacks strength for this sum rule, which
may be caused by the importance of the neglected pionic contributions.
Starting from $Q^2\approx0.2$ GeV$^2$, beyond which region these contributions
are less important, the HCQM is in notable agreement with MAID. \\
The situation is quite different for the neutron (right panel). In this case,
MAID yields a large positive value at $Q^2=0$, while the sum rule requires
$I_{LT}^n(0)\,=\,0$.
On the contrary, the HCQM result is in very good agreement
with the sum rule, whereas the HO model gives a small positive value.
At finite values of momentum transfer, the new data on $g_1^n+g_2^n$
shed more light on the situation with this sum rule, supporting the MAID
predictions in the range $Q^2>0.2$ GeV$^2$. However, an inclusion of the
non-resonant contributions from energy range $W>2$ GeV is necessary to match
these data with the model-independent sum rule value and reduce the
systematic errors (not displayed in Fig. \ref{fig:i3}).\\

\begin{figure}[h]
\begin{center}
\epsfxsize=27pc
\centerline{\epsffile{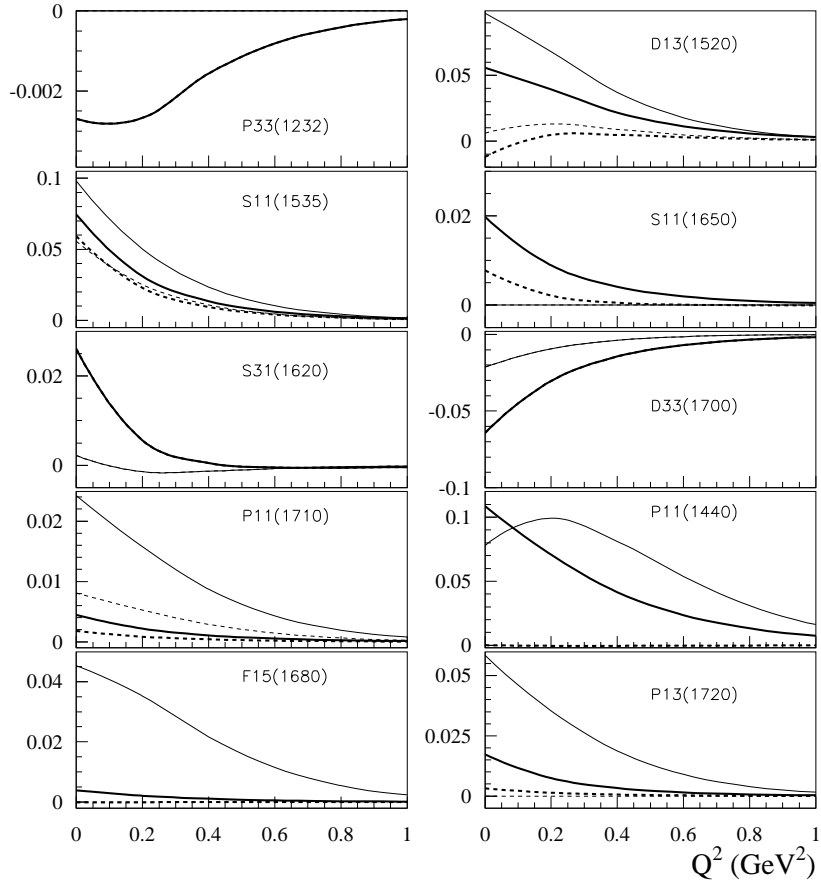}}
\end{center}
\vspace{-1 cm}
\caption{The single resonance contributions to the $I_{LT}(Q^2)$ integral as
predicted by the HCQM for the proton (thick full lines) and for the neutron
(thick dashed lines). The HO results are shown by the respective thin lines.}
\label{fig:i3rescontr}
\end{figure}

The vanishing of $I_{LT}^n(0)$ requires, of course, an exact cancellation of all
the resonance and background contributions. In order to track the mechanism of
cancellations among the different
resonances, the various contributions are separately displayed in
Fig. \ref{fig:i3rescontr}, for both proton (solid lines) and neutron
(dashed lines). 

As can be clearly seen, the mechanism of this cancellation is
quite different in the two models. In the case of HCQM, the major
contributions are due to the
resonances $D_{13}(1520)$, $P_{11}(1440)$, $S_{11}(1535)$, and $D_{33}(1700)$.
The latter two cancel for both the proton and the neutron case. On the other
hand, the contributions of the resonances $D_{13}(1520)$ and $P_{11}(1440)$
are rather large for the proton but practically zero in the neutron case,
which leads to a small sum rule value for the neutron.
The presented HO calculation does not
contain the $SU(6)$-breaking term, therefore most of the resonance
contributions are zero in the case of the neutron.
In the case of MAID, the failure
to reproduce this zero may be related to the neglect of two-pion and heavier
mass final states. However, it is even more likely that the discrepancy is
due to the neglect of the spectrum above the $W=2$~GeV, because $I_{LT}$
clearly has the worst convergence of all the integrals.

\subsection{Longitudinal-transverse polarizability $\delta_{LT}$}

\begin{figure}[h]
\begin{center}
\epsfxsize=27pc
\centerline{\epsffile{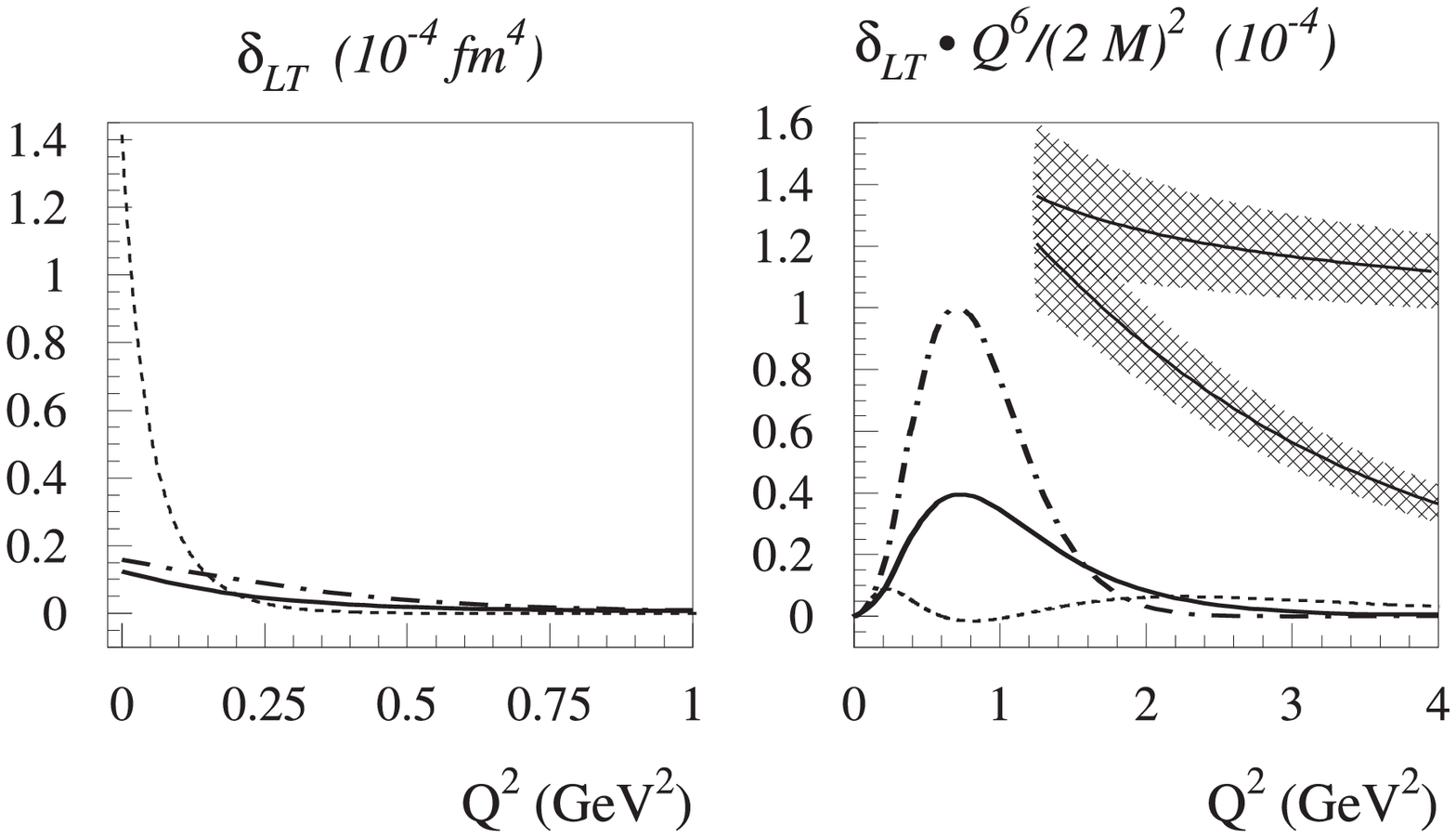}}
\end{center}
\vspace{-1 cm}
\caption{Results for $\delta^p_{LT}$ (left panel) in units of $10^{-4}$ fm$^4$,
and $\delta_{LT}\cdot Q^6/(2M)^2$ in units of $10^{-4}$. For further notation
see Fig.\ref{fig:gamma0}.}
\label{fig:deltalt}
\end{figure}

We now discuss the polarizability $\delta_{LT}$ involving the
longitudinal-transverse interference term $\sigma_{LT}$ for the case of the
proton. Figure \ref{fig:deltalt} displays the predictions of the two CQMs in
comparison with MAID. The latter predicts a large positive value for
$\delta_{LT}$ at $Q^2=0$ due to the interference term $E_{0+}\cdot S^*_{0+}$,
dominated by the near threshold $s$-wave pion production but absent in the CQM.
 However, starting from $Q^2\approx0.2$ GeV$^2$, the pionic contribution
becomes small, and all the models give similar results. The right panel of Fig.
\ref{fig:deltalt} allows for a direct comparison to the integral evaluated with
the DIS structure functions $g_1$ and $g_2$. As we can see, neither of the
models matches to the DIS data, however the CQM models seem to work better in
the range of $0.2$ GeV$^2<Q^2<2$ GeV$^2$.

\subsection{Longitudinal polarizability $\alpha_{L}$}

\begin{figure}[h]
\begin{center}
\epsfxsize=27pc
\centerline{\epsffile{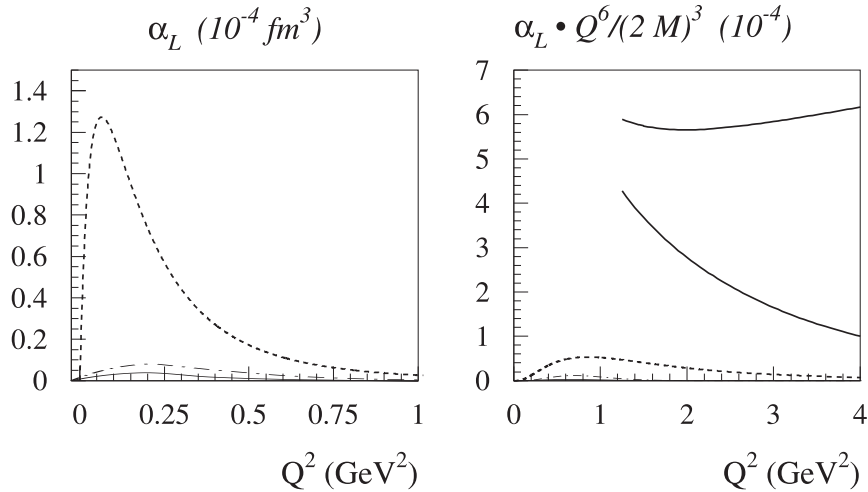}}
\end{center}
\vspace{-1 cm} \caption{Results for $\alpha_L$ (left panel) in units of
$10^{-4}$ fm$^3$, and $\alpha_L\cdot Q^6/(2M)^3$ in units of $10^{-4}$. For
further notation see Fig.\ref{fig:gamma0}.}
\label{fig:alphal}
\end{figure}

We finally show our results for the longitudinal polarizability $\alpha_L$ for
the proton, in order to test our model for the purely longitudinal transitions.
As becomes evident from Fig.\ref{fig:alphal}, all the presented models are
incompatible with the DIS data over the full range covered by these data. This
indicates that there is still space for further theoretical investigations of
the longitudinal photon coupling to the nucleon.

\section{Conclusions}
We have presented a calculation of the generalized sum rules and
polarizabilities of the nucleon in the non-relativistic constituent quark model
with the hypercentral potential. This model has been used previously to
calculate various observables: the baryonic spectrum \cite{hccqm1}, the
photocouplings \cite{hccqm2}, electromagnetic transition amplitudes
\cite{hccqm3}, elastic nucleon form factors \cite{hccqm4} and the longitudinal
electromagnetic transition form factors \cite{hcqmlong}. The aim of these
calculations was to consistently describe diverse observables within the same
model, with all the parameters fixed to the spectrum, and the present work on
the generalized sum rules is a further step in this direction. The generalized
sum rules provide model-independent relations between the static and dynamic
properties of the nucleon, and therefore only a model which correctly accounts
for all the relevant mechanisms can be able to obey these sum rules. In the
previous CQM calculations of the GDH sum rule \cite{gdhcqm} the emphasis was on
the reproduction of the sum rule value at $Q^2=0$ which amounts to relating the
anomalous magnetic moment of the nucleon to its excitation spectrum. However,
in the last few years a lot of experimental data have become available on the
GDH and related sum rules over a wide range of $Q^2$, which allows us to test
our model as function of the momentum transfer. In turn, this corresponds to
testing the spatial distribution of the quark charge, spin, and momentum inside
the nucleon. Since the CQM operates only with the heavy constituent quark
degrees of freedom and neglects the quark-antiquark sea effects, it is expected
that the model should experience difficulties at low $Q^2$ values, where these
neglected effects become dominant. With the possible exception of
$\gamma_{TT}^p$ and $I_{LT}^n(0)$, our results clearly show that the CQM does
not adequately describe the discussed integrals in the range $0<Q^2<0.2$
GeV$^2$. We therefore conclude that the simple CQM description for the response
of the nucleon to a quasistatic electromagnetic field is not adequate, and that
an inclusion of the pionic ($q\bar{q}$) contributions is crucial in this
kinematical region.

Such ``pion cloud'' effects are known to drop very fast with $Q^2$, and beyond
$Q^2\approx0.2$ GeV$^2$, the HCQM agrees qualitatively with the corresponding
integrals evaluated with the DIS structure function over the resonance region
$W\leq2$ GeV$^2$. For comparison, we plotted our predictions along with the
predictions of the harmonic oscillator CQM and the phenomenological MAID model.
The HO model gives very similar results as those of the HCQM at low $Q^2$,
however it drops significantly faster with $Q^2$ due to its characteristic
Gaussian form factors of the HO model, such that the observables are
practically vanishing for $Q^2\geq1$ GeV$^2$. The MAID model accounts for all
the leading mechanisms of pion production, such as resonant and non-resonant
photo- and electroproduction, vector meson exchange, and uses extensive data
sets to fix its parameters. For the proton, MAID works quite well at low and
intermediate momentum transfers, but falls short of the data at the higher
$Q^2$ values, which indicates increasing importance of the contributions above
the resonance region. In the range of $0.2$ GeV$^2<Q^2<2$ GeV$^2$, we found
that the HCQM model predicts most of the proton observables very similar to
MAID. However, MAID fails to reproduce the neutron sum rules at the smaller
momentum transfer, whereas the HCQM works for quite well in this case,
especially for $I_{LT}(Q^2)$ and $I_{TT}(Q^2)$. \\

This work was partially supported by INFN and MIUR (M.G., M.M.G. and E.S.), and
by the Deutsche Forschungsgemeinschaft (D.D. and L.T.).

\begin{appendix}
\section{Helicity representation of the inclusive cross section}
\label{sec:appendix}
In this appendix we derive the inclusive cross section of Eq.(1) for the
scattering of an electron with definite helicity off a polarized nucleon
characterized by its spin projection or helicity. Since we neglect the mass
of the electron, its helicity $h$ cannot change in the scattering process.
The hadronic transition leads from the nucleon with helicity $\Lambda$ to an
excited state with spin $J$ and helicity $\Lambda'$. We note that all
helicities are defined as projections of the spin on the direction of the
virtual photon. The invariant matrix element for such a transition takes the
form
\beqn
T_{\Lambda' h'\Lambda h} &=& \frac{e^2}{Q^2}\bar{u}(E',h')
\gamma_\mu u(E,h)\, <J\Lambda'|J^\mu|\textstyle{\frac{1}{2}}\Lambda>\nn\\
&=&\frac{e^2}{Q^2} \sum_\lambda\Omega_{h,\lambda}\delta_{h'h}
<J\Lambda'|\varepsilon^{(\lambda)}\cdot J|\textstyle{\frac{1}{2}}\Lambda>\;,
\label{A1}
\eeqn
where we inserted the complete set of polarization vectors of Eq.~(20). The
leptonic matrix element for a given helicity is
\beqn
\Omega_{h,\lambda}\;=\; \frac{\bar{u}(E',h')\varepsilon^{(\lambda)*}
\cdot\gamma\,
u(E,h)} {\varepsilon^{(\lambda)*}\cdot \varepsilon^{(\lambda)}}\;.
\eeqn

We evaluate this matrix element in the $lab$ frame, choosing the x-z plane as
the leptonic plane and with the virtual photon momentum $\vec{q}$ pointing in
the direction of the $z$ axis. The four-momenta of the electron are then given
by
\beqn
k^\mu &=& E(1,\sin\Theta_1,0,\cos\Theta_1)\nn\ ,\\
k'^\mu &=& E'(1,\sin\Theta_2,0,\cos\Theta_2)\ ,
\eeqn
with $E$ and $E'$ the initial and final electron energy, respectively,
$\Theta_1$ and
$\Theta_2$ the corresponding polar angles, and $\Theta=\Theta_2-\Theta_1$ the
scattering angle. The virtual photon has the four-momentum
\beqn
q^{\mu} = (\nu,0,0,|\vec{q}|)\ ,
\eeqn
where $\nu=E-E'$ and $Q^2=\vec{q}\,^2-\nu^2>0$.

The following kinematical relations are useful for the further calculation:
\beqn
E\sin\Theta_1&=&E'\sin\Theta_2\;=\;\frac{EE'\sin\Theta}{q}\;,\nn\\
\cos\Theta_1&=&\frac{E-E'\cos\Theta}{q}\;,\nn\\
\cos\Theta_2&=&\frac{E\cos\Theta-E'}{q}\;,\nn\\
\eeqn

The initial state electron spinor has the form
\beqn
u(E,h)\;=\;\sqrt{E} \left( \begin{array}{c} 1
\\ h \end{array}  \right)\ \chi_h \ ,
\eeqn
where $\chi_h$ is a Pauli spinor,
\beqn
\chi_{+} &=& \left( \begin{array}{c} \cos{\Theta_1\over 2}\\
\sin{\Theta_1\over 2} \end{array} \right) \;\;\;\;\;
\chi_{-}\;=\; \left( \begin{array}{c} -\sin{\Theta_1\over 2}\\
\cos{\Theta_1\over 2} \end{array} \right)\ .
\eeqn

The final state spinor is then obtained by the replacement $E\rightarrow E'$,
$h\rightarrow h'$, and $\Theta_1\rightarrow \Theta_2$. We further note that
the ``helicities'' $h$ und $h'$, as defined in Eq.~(1), have the values
$\pm1$. With these definitions the leptonic matrix element takes the form:
\beqn
\Omega_{h,\lambda}\;=\; \left\{ -\frac{\lambda}{\sqrt{2}}
\left[ \sqrt{\frac{1+\epsilon}{1-\epsilon}} + h\lambda \right] \,+\,
\sqrt{\frac{2\epsilon}{1-\epsilon}}\delta_{\lambda 0} \right\}\,Q\ .
\label{A8}
\eeqn

In deriving this equation, the following relations are useful:

\beqn
\sin\frac{\Theta}{2} & = & \frac{Q}{2\sqrt{E'E}}\;,\nn\\
\cos\frac{\Theta}{2} & = & \sqrt{\frac{2\epsilon}{1-\epsilon}} \
\frac{q}{Q}\sin\frac{\Theta}{2}\;,\nn \\
\sin\left(\frac{\Theta_1+\Theta_2}{2}\right)
&=& \sqrt{\frac{1+\epsilon}{1-\epsilon}}\,\sin\frac{\Theta}{2}\;.
\eeqn

We now return to the hadronic matrix element in Eq.~(\ref{A1}) and assume
that the target nucleon is fully polarized in the direction
\beqn
\hat{P}
= (\sin\Psi\cos\Phi,\,\sin\Psi\sin\Phi,\,\cos\Psi) = (P_x,\,P_y,\,P_z)\ .
\eeqn

In general the nucleon is therefore in a superposition of states with definite
helicity,
\beqn
|N\rangle\;=\;\cos\textstyle{\frac{\Psi}{2}}|\textstyle{\frac{1}{2}}\,,
\textstyle{\frac{1}{2}}\rangle
\,+\,
e^{i\Phi}\sin\textstyle{\frac{\Psi}{2}}|\textstyle{\frac{1}{2}}\,, -
\textstyle{\frac{1}{2}}\rangle \ .
\eeqn

Equation (\ref{A1}) involves transitions for all values of
$\lambda,\ \Lambda,$ and $\Lambda' = \lambda+\Lambda$, which gives 6 different
matrix elements. However, the helicity amplitudes obey the relation
\beqn
\langle J,-\Lambda'|\varepsilon^{(-\lambda)}
\cdot J|\textstyle{\frac{1}{2}},-\Lambda\rangle
\;=\;
\xi\langle J,\Lambda'|\varepsilon^{(\lambda)}
\cdot J|\textstyle{\frac{1}{2}},\Lambda\rangle\,,
\eeqn
i.e., the sign reversal of all the involved helicities only leads to an
overall phase,
$|\xi|=1$. Moreover, with an appropriate definition of the wave functions,
all helicity amplitudes turn out to be real, and therefore $\xi = \pm 1$.
While the phase $\xi$ depends on the quark wave function of the respective
resonance, it has the same value for all 3 components of the current. We can
therefore relate all hadronic matrix elements to the 3 helicity amplitudes
of Eq.~(22) and the phase $\xi$.
Equation~(\ref{A1}) can now be cast
into the form:
\beqn
T_{\Lambda' h'\Lambda h} & \sim &
\Omega_{h,0}
\left(
\cos\textstyle{\frac{\Psi}{2}}
\,\delta_{\Lambda',\frac{1}{2}} \delta_{\Lambda,\frac{1}{2}}
\,+\,
e^{i\Phi}\sin\textstyle{\frac{\Psi}{2}}\,
\delta_{\Lambda',-\frac{1}{2}}
\delta_{\Lambda,-\frac{1}{2}}
\,\xi
\right )
\frac{Q}{q} S_{\frac{1}{2}} \nn \\
&+&
\Omega_{h,1}
\left(
\cos\textstyle{\frac{\Psi}{2}}\, A_{\frac{3}{2}}
\delta_{\Lambda',\frac{3}{2}}\delta_{\Lambda,\frac{1}{2}}
\, +\,
e^{i\Phi}\sin\textstyle{\frac{\Psi}{2}}\,
A_{\frac{1}{2}} \delta_{\Lambda',\frac{1}{2}}\delta_{\Lambda,-\frac{1}{2}}
\right ) \\
&+& \Omega_{h,-1} \left( \cos\textstyle{\frac{\Psi}{2}}\, \xi\,A_{\frac{1}{2}}
\delta_{\Lambda',-\frac{1}{2}} \delta_{\Lambda,\frac{1}{2}} \,+\,
e^{i\Phi}\sin\textstyle{\frac{\Psi}{2}}\, \xi\, A_{\frac{3}{2}}
\delta_{\Lambda',-\frac{3}{2}} \delta_{\Lambda,-\frac{1}{2}} \right ).\nn \eeqn

In order to derive the cross section, we sum the absolute squares of the matrix
elements over the final state helicities: \beqn
\sum_{h',\Lambda'}|T_{h'\Lambda' h\Lambda}|^2 & \sim & |\Omega_{h,0}
\cos\textstyle{\frac{\Psi}{2}}\, \frac{Q}{q} S_{\frac{1}{2}} + \Omega_{h,1}
e^{i\Phi} \sin\textstyle{\frac{\Psi}{2}}\, A_{\frac{1}{2}}|^2
\nn\\
&+&
|\Omega_{h,0} e^{i\Phi} \sin\textstyle{\frac{\Psi}{2}}\,
 \xi \frac{Q}{q} S_{\frac{1}{2}} +
\Omega_{h,-1} \cos\textstyle{\frac{\Psi}{2}}\, \xi A_{\frac{1}{2}}|^2
\nn\\
&+&
| \Omega_{h,1} \cos\textstyle{\frac{\Psi}{2}}\, A_{\frac{3}{2}}|^2
+ | \Omega_{h,-1} e^{i\Phi} \sin\textstyle{\frac{\Psi}{2}}\, \xi
A_{\frac{3}{2}}|^2 \ .
\label{A14}
\eeqn

It is now evident that the state-dependent phase $\xi$ drops out in the
inclusive cross section. Therefore, the above equation can be cast into the
form:
 \beqn \sum_{h',\Lambda'}|T_{h'\Lambda' h\Lambda}|^2 & \sim &
 \epsilon\frac{Q^2}{q^2}\,|S_{\frac{1}{2}}|^2
\,+\,
\frac{1}{2} (1-\cos\Psi\sqrt{1-\epsilon^2})\,|A_{\frac{1}{2}}|^2 \nn \\
&+&
\frac{1}{2} (1+\cos\Psi\sqrt{1-\epsilon^2})\,|A_{\frac{3}{2}}|^2  \nn \\
&-&
\sin\Psi\cos\Phi h\sqrt{2\epsilon (1-\epsilon)}
\frac{Q}{\sqrt{2}\,q}\,{\rm Re}(S^*_{\frac{1}{2}}A_{\frac{1}{2}}) \nn\\
&-&
\sin\Psi\sin\Phi \sqrt{2\epsilon (1+\epsilon)}
\frac{Q}{\sqrt{2}\,q}\,{\rm Im}(S^*_{\frac{1}{2}}A_{\frac{1}{2}})\,.
\eeqn

The last term in this equation must vanish in the inclusive cross section, see
Eq.(1). This requires that $S^*_{\frac{1}{2}}A_{\frac{1}{2}}$ is a real number,
which is of course also the case in any quark model calculation. In the
following we shall therefore treat the helicity amplitudes as real numbers. Our
final result takes the following form: \beqn \sum_{h',\Lambda'}|T_{h'\Lambda'
h\Lambda}|^2 & \sim & \frac{1}{2} (A_{\frac{1}{2}}^2+A_{\frac{3}{2}}^2) \,+\,
\epsilon \frac{Q^2}{q^2}\,S_{\frac{1}{2}}^2\nn\\
&-&
h\,P_z\sqrt{1-\epsilon^2} \,
\frac{1}{2} (A_{\frac{1}{2}}^2-A_{\frac{3}{2}}^2)  \nn\\
&-&h\,P_x \sqrt{2\epsilon (1-\epsilon)} \frac{Q}{\sqrt{2}\,q}\,S_{\frac{1}{2}}
A_{\frac{1}{2}}\ , \eeqn with $P_z = \cos\Psi, P_x=\sin\Psi\cos\Phi$, and
independent of $P_y$. Comparing finally with Eq.(1) we find, up to a common
factor, the results of Eq.(27), \beqn \sigma_T \sim \frac{1}{2}
(A_{\frac{1}{2}}^2+A_{\frac{3}{2}}^2) \quad,\quad
\sigma_{TT} \sim \frac{1}{2} (A_{\frac{1}{2}}^2-A_{\frac{3}{2}}^2)\ ,\nn \\
\sigma_L \sim \frac{Q^2}{q^2}\,S_{\frac{1}{2}}^2 \quad,\quad \sigma_{LT} \sim
-\frac{Q}{\sqrt{2}\,q}\, S_{\frac{1}{2}} A_{\frac{1}{2}}\ . \eeqn

\end{appendix}

\end{document}